\documentclass[conference]{IEEEtran}
\IEEEoverridecommandlockouts
\usepackage{cite}
\newcommand{\e}[1]{\mbox{\lstinline[basicstyle=\small]|#1|}}

\usepackage{boogie}
\usepackage{eiffel}
\usepackage{listings}
\usepackage{amsmath,amssymb,amsfonts}
\usepackage{algorithmic}
\usepackage{graphicx}
\usepackage{textcomp}
\usepackage{xcolor}
\usepackage[linesnumbered,ruled,vlined]{algorithm2e}

\usepackage[utf8]{inputenc}

\newcommand{\tabincell}[2]{\begin{tabular}{@{}#1@{}}#2\end{tabular}}

\makeatletter
\algocf@newcmdside@kobe{FromLoop@from}{%
    \KwSty{from}%
    \ifArgumentEmpty{#1}\relax{ #1}%
    \algocf@group{#2}%
    \par
}
\algocf@newcmdside@kobe{FromLoop@until}{%
    \KwSty{until}%
    \ifArgumentEmpty{#1}\relax{ #1}%
    \algocf@group{#2}%
    \par
}
\algocf@newcmdside@kobe{FromLoop@loop}{%
    \KwSty{loop}%
    \ifArgumentEmpty{#1}\relax{ #1}%
    \algocf@block{#2}{end}{#3}%
    \par
}

\newcommand\FromLoop[3]{%
    \FromLoop@from{#1}%
    \FromLoop@until{#2}%
    \FromLoop@loop{#3}%
}

\newcommand\IfThenElse[2]{%
    \IfThenElse@if{#1}%
    \IfThenElse@elseif{#2}%
}

\algocf@newcmdside@kobe{IfThenElse@if}{%
    \KwSty{if}%
    \ifArgumentEmpty{#1}\relax{ #1}  
    \algocf@group{#2}
    \par
}
\algocf@newcmdside@kobe{IfThenElse@elseif}{%
    \KwSty{elseif}%
    \ifArgumentEmpty{#1}\relax{ #1}%
    \algocf@group{#2}%
    \par
}

\def\BibTeX{{\rm B\kern-.05em{\sc i\kern-.025em b}\kern-.08em
    T\kern-.1667em\lower.7ex\hbox{E}\kern-.125emX}}
\begin{document}

\title{Improve Counterexample Quality}
\title{Improve Counterexample Quality for Program Verification}
\title{Improve Counterexample Quality for Eiffel Program Verifier}
\title{Improving Counterexample Quality from Failed Program Verification}

\author{\IEEEauthorblockN{Li Huang}
\IEEEauthorblockA{\textit{Chair of Software Engineering} \\
\textit{Schaffhausen Institute of Technology}\\
Schaffhausen, Switzerland \\
li.huang@sit.org}
\and
\IEEEauthorblockN{Bertrand Meyer}
\IEEEauthorblockA{\textit{Chair of Software Engineering} \\
\textit{Schaffhausen Institute of Technology}\\
Schaffhausen, Switzerland\\
bm@sit.org}
\and
\IEEEauthorblockN{Manuel Oriol}
\IEEEauthorblockA{\textit{Chair of Quantum Software Engineering} \\
\textit{Schaffhausen Institute of Technology}\\
Schaffhausen, Switzerland\\
mo@sit.org}
}

\maketitle

\begin{abstract}
In software verification, a successful automated program proof is the ultimate triumph. 
The road to such success is, however, paved with many failed proof attempts. 
The message produced by the prover when a proof fails is often obscure, making it very hard to know how to proceed further.
The work reported here attempts to help in such cases by providing immediately understandable counterexamples. 
To this end, it introduces an approach called Counterexample Extraction and Minimization (CEAM). When a proof fails, CEAM turns the counterexample model generated by the prover into a a clearly understandable version; it can in addition simplify the counterexamples further by minimizing the integer values they contain. We have implemented the CEAM approach as an extension to the AutoProof verifier and demonstrate its application to a collection of examples.
\end{abstract}


\section{Introduction}
\noindent Deductive program verification performs a rigorous analysis of the correctness of programs with respect to their functional behavior, usually specified formally by \emph{contracts} (such as pre- and postconditions, can class and loop invariants). The approached has progressed in recent years thanks to the
 development of powerful proof engines. 
In practice, however, verifying industrial applications remains difficult. 
One of the obstacles is the lack of intuitive feedback to understand the reasons why a verification attempt failed.  
Although in many cases the underlying prover can provide a counterexample containing some usable diagnostic information for debugging, such a counterexample usually contains hundreds of difficult-to-interpret lines. Another obstacle to usability is that integer values generated by the prover for the counterexamples are often very large, and hence do not provide programmers with an easy intuitive understanding of what is wrong.

This article presents the Counterexample Extraction and Minimization (CEAM) approach for improving the quality of counterexamples produced when a proof fails, and making them usable for identifying and correcting the underlying bugs.
We have implemented CEAM as an extension of the AutoProof environment~\cite{autoproof, tschannen2015autoproof}, a static verification platform for contract-equipped Eiffel~\cite{meyer1997object} programs based on Hoare-style proofs. AutoProof relies on the Boogie proof system~\cite{barnett2005boogie, le2011boogie} and takes advantage of Boogie's underlying SMT (Satisfiability Modulo Theories) solver, by default Z3 \cite{de2008z3}.  

When a proof fails, CEAM exploits the counterexample models (hereinafter referred to simply as models) generated by the SMT solver   and generates simple counterexamples in a format more intuitive to programmers. 
CEAM also provides a minimization mechanism allowing programmers to get simplified counterexamples with integer variables reduced to their minimal possible values. The current version of CEAM supports primitive types (integer, boolean), user-defined types (classes) as well as some commonly used container types (such as arrays and sequences).

Section \ref{sec: example} illustrates an example of using the CEAM approach. Section \ref{sec: technology_stack} introduces the technologies used in our verification framework. Section \ref{sec: implementation} describes the details of the implementation of the CEAM. 
We evaluate the applicability of the CEAM through a series of examples in Section \ref{sec: experiment}. 
After a review of related work in Section \ref{sec: related_work}, Section \ref{sec: conclusion} concludes the paper with our ongoing work.

\section{An Example Session}
\label{sec: example}
Before exploring the principles and technologies of the CEAM approach, we look at its practical use on a representative example (Fig.~\ref{listing: max}). The intent of the \e{max} function in class \e{MAX} is to return into \e{Result} the maximum element of an integer array \e{a} of size \e{a.count}. The two postconditions in lines 22 and 23 (labeled by \emph{is\_max} and \emph{in\_array}) specify this intent: every element of the array should be less than or equal to \e{Result}; at least one element should be equal to \e{Result}.

\begin{figure}[htbp]
\centerline{{\includegraphics[width=3.5in]{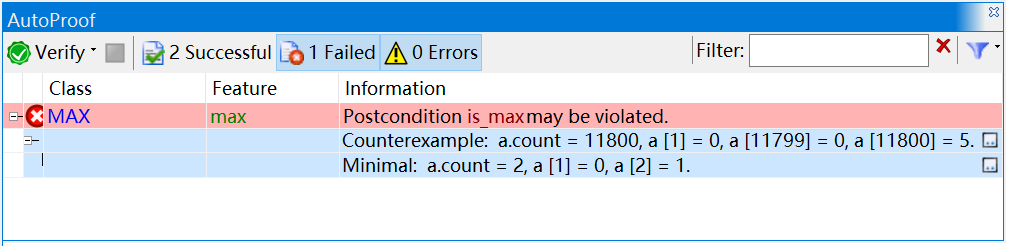}
}}
\vspace{-0.1in}
\caption{Proof result in AutoProof: the first row (highlighted in red) indicates a proof failure; the second row is a counterexample generated based on the Z3 model; the third row is a minimized counterexample.}
\label{fig: proof-result}
\end{figure}

When we try to verify the \e{max} function using AutoProof, verification fails and AutoProof returns an error message ``Postcondition \emph{is\_max} may be violated'' (the first row in Fig. \ref{fig: proof-result}). 
Such a generic message tells us that the prover cannot establish the postcondition, but does not enable us to find out why. In this case, programmers can look at the model generated by the Z3 solver to understand the cause of the failure. Deciphering the model is a cumbersome task: the model spans hundreds of lines and is expressed in a cryptic formalism.

In contrast, AutoProof extended with CEAM automatically generates a much simpler counterexample from the model. As displayed in the second row of Fig.~\ref{fig: proof-result}, the counterexample contains the initial values (on entry of \e{max}) of the array's size and of some of its elements. Seeing these concrete values, rather than just the prover's general failure message, helps the programmer conjecture possible reasons for the failure. 
The values in the counterexample are  large, however, too large to give the programmer a direct intuition of the problem at a human scale. 

\vspace{-0.2in}
\begin{figure}[htbp]
   \centering
\begin{lstlisting}[captionpos=b, basicstyle=\fontsize{0.25cm}{0.25cm}]
class MAX feature
max (a: ARRAY $[$INTEGER$]$): INTEGER	
  require   a.count $>$ $0$ 
  local	  i: INTEGER
  do
      from 
          Result $:=$ a $[1]$; i $:=$ $2$
      invariant
          $2$ <= i and i <= a.count $+$ $1$
          $\forall$ j: $1$ $|..|$ $($i - $1)$ | a.sequence $[$j$]$ $<=$ Result
          $\exists$ j: $1$ $|..|$ $($i - $1)$ | a.sequence $[$j$]$ $=$ Result
      until
          i $\geq$ a.count
      loop
          if a $[$i$]$ > Result then
              Result $:=$ a $[$i$]$
          end
          i $:=$ i $+$ $1$
      variant   a.count - i
      end
   ensure
      is_max:   $\forall$ j: $1$ $|..|$ a.count | a.sequence $[$j$]$ $<=$ Result 
      in_array: $\exists$ j: $1$ $|..|$ a.count | a.sequence $[$j$]$ $=$ Result 
   end
end
\end{lstlisting}
\vspace{-0.15in}
\caption{{\tiny\e{MAX}} is a class that finds the maximum element of an integer array; a fault (the exit condition at line 13 is incorrect) is injected to the code for presentation purposes.}
\label{listing: max}
\end{figure}  

\vspace{-0.1in}

\noindent To provide a more intuitive illustration, CEAM allows the programmer to query AutoProof further to obtain a \textit{minimal} counterexample in the last row of Fig.~\ref{fig: proof-result}, where integer variables have been reduced to their minimal possible values. 
\vspace{-0.1in}

\noindent This minimized counterexample provides a simple diagnostic trace of \e{max}: on loop initialization at line 7, \e{Result} = 0 and i = 2; at line 13, the exit condition of the loop evaluates to \e{True} with \e{a.count} = 2 and \e{i} = 2, which forces the loop to terminate. These values reveal the fault in the program: the loop terminates too early, preventing the program from getting to the actual maximum value, found at position 2 of the array \e{a}. To eliminate this error, it suffices to strengthen the exit condition to permit one more loop iteration: change \e{i} \e{$\geq$} \e{a.count} to \e{i} \e{$>$} \e{a.count}.





\section{Technology stack}
\label{sec: technology_stack}
\noindent This section introduces technologies used in the present work, including language and prover.

\textbf{Eiffel}~\cite{meyer1997object, bertrand2016touch} is an object-oriented programming language which natively supports Design-by-Contract\cite{meyer1992applying}. An Eiffel program consists of a set of \emph{classes}. A class represents a set of run-time objects characterized by the \textit{features} applicable to them. 
Fig. \ref{listing: ACCOUNT} shows a simple class representing bank accounts. The class contains two types of features: 
\emph{attributes} representing data items associated with instances of the class, such as \e{balance} (line 2) and \e{credit\_limit} (line 4); 
\emph{routines} representing operations applicable to these instances, including \e{available\_amount} and \e{transfer}. 
Routines are further divided into \emph{procedures} (with no returned value) and \emph{functions} (returning a value). Here, \e{available\_amount} is a function returning an integer (represented by the special variable \e{Result}), and \e{transfer} is a procedure. 

\vspace{-0.2in}
\begin{figure}[htbp]
\centering
\begin{minipage}{.43\textwidth}
\begin{lstlisting}[basicstyle=\linespread{0.1}\fontsize{0.27cm}{0.27cm}]
class ACCOUNT feature
balance: INTEGER 
    -- Balance of this account.
credit_limit: INTEGER
    -- Credit limit of this account.
available_amount: INTEGER	
    -- Amount available on this account.
  do
      Result $:=$ balance - credit_limit
  end
transfer (amount: INTEGER; other: ACCOUNT)	
  -- Transfer `amount' to the `other' account.
  require
      amount $>=$ $0$ 
      amount $<=$ available_amount
  do
      balance $:=$ balance - amount
      other.balance $:=$ other.balance $+$ amount
  ensure
      withdrawal: balance $=$ old balance - amount
      deposit: other.balance $=$ old other.balance $+$ amount
  end
end
\end{lstlisting}
\end{minipage}
\vspace{-0.1in}
\caption{A class implementing the behavior of bank accounts}
    \label{listing: ACCOUNT}
\end{figure}

\vspace{-0.1in}
\noindent Programmers can specify the properties of Eiffel classes by equipping them with contracts of the following types:
\begin{itemize} 
\item A precondition (\e{require}) must be satisfied at the time of any call to the  routine;  the precondition of \e{transfer} (lines 13 -- 15), for example, requires the value of \e{amount} to be non-negative an no greater than \e{available_amount}.  
\item A postcondition (\e{ensure}) must be guaranteed on routine's exit; for instance, a postcondition of \e{transfer} at line 20 states that, at the end of the execution of \e{transfer}, the value of \e{balance} must have been decreased by \e{amount}.
\item A loop invariant (\e{invariant}) characterizes the semantics of a loop in the form of a property satisfied after initialization and preserved by every iteration, as illustrated by the invariant of \e{max} (lines 9 -- 11 in Fig. \ref{listing: max}) specifies the properties of \e{i} and \e{Result} before and after every iteration.  
\item A loop variant (\e{variant}) is an integer measure that should always be non-negative and decrease strictly at each loop iteration, ensuring that the loop eventually terminates; the loop variant of the loop in \e{max} is \e{a.count - i} (line 19 in Fig. \ref{listing: max}).
\end{itemize}

\noindent Contracts embedded in the code make it amenable to both dynamic analysis (run-time checking of the contract properties), as in the EiffelStudio environment, and static analysis (Hoare-style proofs), as in AutoProof.

\vspace{0.05in}
\textbf{AutoProof} \cite{autoproof, tschannen2015autoproof} is a static verifier that checks the correctness of Eiffel programs against their functional specifications (contracts). 

When verifying an Eiffel program, AutoProof translates the program into a Boogie program  \cite{barnett2005boogie, le2011boogie}, which is then transformed into a set of verification conditions (VCs) in SMT-LIB \cite{barrett2010smt} format, based on Dijkstra's weakest precondition calculus \cite{dijkstra1976discipline}. 
The program's correctness follows from the validity of the VCs. 
Boogie asks an SMT solver (by default Z3, as noted) to reason about the validity of each VC. Specifically, the solver tries to find a model (an interpretation of variables and functions used in the SMT encodings) that satisfies the negation of a VC. 
If the solver is unable to find such a model (no counterexample exists and thus VC is a tautology), the verification is successful.
If it succeeds in obtaining such a model, the verification fails and the solver makes the model available. This model witnesses the invalidity of the VC \cite{leino2005generating} and thus can be seen as a counterexample\footnote{The counterexample is a \textit{potential} counterexample since it can occasionally be spurious because of the prover's incompleteness, although that phenomenon is not significant in our experience.} at the SMT level. In general, an SMT model describes an execution trace (a sequence of program states) of a failed routine, along which the program goes to an error state. The CEAM approach makes use of such models to generate easy-to-understand counterexamples.

\section{Counterexample extraction and minimization}
\label{sec: implementation}
\noindent This section first shows how to generate counterexamples based on SMT models, then presents the details of the CEAM strategy for counterexample minimization. 

\subsection{Counterexample extraction}
\noindent In general, to construct a counterexample it suffices to extract the concrete values of relevant variables from the corresponding SMT model, and to use these values to produce a counterexample message (as in Fig. \ref{fig: proof-result}). 
The format of the message can vary depending on the chosen ``verbosity level''. As the goal of the approach and the tools is to ease the burden on programmers, the message only displays the initial  values of relevant input variables in the counterexample, as illustrated in the case at the beginning of this article.

Fig. \ref{listing: withdrawal_counterexample} shows a simplified portion of the Z3 model\footnote{As the models are too big to include in their entirety, this presentation only displays the parts relevant to the discussion.} corresponding to the failed proof of \e{transfer}'s postcondition labeled by \emph{withdrawal} (line 20 in Fig. \ref{listing: ACCOUNT}).
To construct a counterexample for this failure, it suffices to obtain the initial values of its three input variables: the implicit variable \e{Current}\footnote{{\ttfamily{Current}} represents the active object in the current execution context, similar to {\ttfamily{this}} in Java.} and the two arguments \e{amount} and \e{other}. 

\vspace{-0.2in}
\begin{figure}[htbp]
    \centering
\begin{minipage}{.43\textwidth}
\begin{lstlisting}[language = Java, basicstyle=\fontsize{0.27cm}{0.27cm}]
amount -> $5799$
Heap -> T@U!val!$17$
Current -> T@U!val!$18$
other -> T@U!val!$18$
ACCOUNT.balance -> T@U!val!$7$
ACCOUNT.credit_limit -> T@U!val!$8$
Select -> {
T@U!val!$17$ T@U!val!$18$ T@U!val!$7$ -> $(- 2147475928)$
T@U!val!$17$ T@U!val!$18$ T@U!val!$8$ -> $(- 2147481727)$
}
\end{lstlisting}
\vspace{-0.1in}
\end{minipage}
\caption{A slice of model of the proof failure of \emph{withdrawal}}
\label{listing: withdrawal_counterexample}
\end{figure}

\vspace{-0.1in}
\noindent In the transformation from Eiffel program to SMT code, to encode the execution semantics of an object-oriented program, the evolution of the \emph{heap} (the collection of program objects) during an execution is modeled as a sequence of constants prefixed with \e{Heap}. Here the SMT constant \e{Heap} (line 2) corresponds to the heap at the initial program state of \e{transfer}.
\e{Select} (line 7) is a function for retrieving the values of objects' fields. 
It takes three parameters, i.e., a heap state, an object reference and a data field, and returns the value of the specified field.

As the example shows, the concrete values of primitive variables (such as \e{amount}) are given directly, whereas the values of non-primitive variables (e.g., \e{Current} and \e{other} of \e{ACCOUNT} type) appear in a symbolic form, prefixed with \e{T@U!val!}. Such symbolic values can be seen as abstract memory locations for the corresponding variables (see \cite{leinoboogie, bjorner2018programming}).
The \e{Select} function is available to obtain the values of the corresponding fields.
For example, \e{Current} is an instance of \e{ACCOUNT} and thus has fields \e{balance} and \e{credit\_limit}. The initial value of \e{balance} can be obtained by applying \e{Select} to a tuple made of \e{Heap} = \e{T@U!val!17} (line 2), \e{Current} = \e{T@U!val!18} (line 3) and \e{ACCOUNT.balance} = \e{T@U!val!7} (line 5); the tuple matches the mapping in line 8, therefore the returned value for \e{balance} is {\small $-2147475928$}. Similarly, the value of \e{credit\_limit} can be retrieved through the mapping in line 9. 

To display the value of a non-primitive variable in the resulting counterexample message, the strategy first checks whether the variable has an alias that has been looked up earlier; if so, it displays the alias relation between the variable and its earliest alias in the message; otherwise, it looks up all of its primitive data fields transitively and display them in the message. 

In this example, the model shows that \e{other} and \e{Current} have the same symbolic values. \e{Current} is, consequently, an alias of \e{other}. As \e{Current} is processed prior to \e{other}, the fields of \e{other} will not be looked up and the message will display the alias relation between \e{other} and \e{Current}. 

After applying the above rules, the counterexample can be derived: \e{balance} = {\small $-2147475928$}, \e{credit\_limit} = {\small $-2147481727$},  \e{amount}  = {\small $5799$}, \e{other} = \e{Current}.  

For variables of container types such as arrays or sequences, the resulting message displays the values of their sizes and containing elements. Here, we use the example of \e{max} in Fig. \ref{listing: max} to demonstrate counterexample extraction for container types.

The AutoProof approach specifies container structures in terms of mathematical structures \cite{tschannen2015autoproof}. For example, the content of the input array \e{a} of \e{max} is specified through a special attribute \e{sequence} (see lines 10 -- 11 in Fig. \ref{listing: max}), which represents the mathematical sequence of integer values stored in \e{a}’s cells. 
To obtain the content of \e{a} from the counterexample, we need to get the content of its \e{sequence} field from the model. Fig. \ref{fig: counterexample_2} shows a slice of the model for the proof failure of \e{max}. CEAM first extracts the value of \e{sequence} by querying \e{Select} with the values of \e{Heap}, \e{a}, and {\e{ARRAY\^INTEGER\^.sequence}} (lines 1 -- 3). Those values match to the mapping in line 5, hence the value of \e{sequence} is \e{T@U!val!38}. By using this value, CEAM then queries the two functions \e{Seq\#Length} and \e{Seq\#Item} to get the values of \e{sequence}'s size (line 8) and elements (lines 12 -- 14), respectively.

\vspace{-0.2in}
\begin{figure}[htbp]
    \centering
\begin{minipage}{.45\textwidth}
\begin{lstlisting}[language = Java, captionpos=b, basicstyle=\fontsize{0.27cm}{0.27cm}]
Heap -> T@U!val!$26$
a -> T@U!val!$18$
ARRAY^INTEGER_32^.sequence -> T@U!val!$9$
Select -> {
T@U!val!$26$ T@U!val!$18$ T@U!val!$9$ -> T@U!val!$38$
}
Seq#Length -> {
  T@U!val!$38$ -> $11800$
  T@U!val!$40$ -> $28101$
}
Seq#Item -> {
  T@U!val!$38$ $1$ -> $0$
  T@U!val!$38$ $11799$ -> $0$
  T@U!val!$38$ $11800$ -> $5$
}
\end{lstlisting}
\end{minipage}
\vspace{-0.1in}
\caption{A snippet of the model of proof failure of \emph{is\_max}}
\label{fig: counterexample_2}
\end{figure}

\vspace{-0.1in}
\subsection{Counterexample minimization}
\noindent Some of the extracted values found to cause a failure, such as $-2147481727$ in the above counterexample, are too large to enable a programmer to visualize easily the cause of the proof failure. CEAM can simplify counterexamples by reducing the absolute value of such integers. Program verification is modular, meaning it processes each routine independently; so does minimization.

As the counterexample of a routine $r$ consists of the initial states of its input variables, to minimize a counterexample of $r$ it suffices to minimize each of $r$'s input variables.
The task of minimizing an input variable $x$ can be reduced to a set of integer minimization tasks based on the type of $x$. The procedure minimize\_integer in in Algorithm \ref{alg: minimize variable} minimizes an integer variable. It applies to the absolute value, retaining the sign (lines 1 -- 5). If $x$ is an object reference, the algorithm first checks whether $x$ denotes a container; if yes, it finds the minimal size of $x$ (lines 8 -- 9) and then minimizes $x$'s elements one by one (lines 11 -- 13); otherwise it minimizes each of $x$'s fields (lines 15 -- 16).

\vspace{-0.1in}
\RestyleAlgo{ruled}
\SetKwComment{Comment}{/* }{ */}
\begin{algorithm}[htbp]
\caption{minimize\_general ($x$): minimize a variable of integer or reference type}
\label{alg: minimize variable}
     \uIf{ x is an integer}{
        \uIf{v $>$ 0}{
            minimize\_integer ($x$)
        }
        \KwSty{else}\uIf{v $<$ 0}  {    
            minimize\_integer ($- x$)
        }
        }
     \KwSty{else}\uIf{x is an object reference}{
         \eIf{$x$ is a container}{
              minimize\_integer ($x.count$)
              \\ $n \gets$ minimized value of $x.count$ \\
              \FromLoop{$i \gets 1$}{$i \leq n$}{
                   minimize\_general ($x[i]$)
                   \\ $i \gets i + 1$
                 }
             }
      {
	         \KwSty{across} {each field $f$ of $x$ \KwSty{as} $x.f$} \KwSty{loop}
	          \ \ \ \ \ minimize\_general ($x.f$)
	    }
	 }
\end{algorithm}
\vspace{-0.1in}

\noindent Algorithm \ref{alg: minimize integer} shows the details of minimize\_integer. $B$ represents the Boogie procedure of routine $r$ generated by AutoProof. The core idea is to find the smallest integer bound $m$ such that adding a precondition $0 \leq x < m$ (line 10) to $B$ still yields the same verification results. When the algorithm ends (no smaller value of $m$ can be found), the model from the last verification run is the minimal possible.

\begin{algorithm}[htbp]
\caption{minimize\_integer ($x$): minimize an integer}
\label{alg: minimize integer}
     \FromLoop{
       $m$ $\gets$ current value of $x$
      \\ $B$.add\_precondtion ($0 \leq x$ $\wedge$ $x < m$)
      \\ verify
     }
     {$no\ smaller\ value\ can\ be\ tried$}
     {
     $B$.remove\_last\_precondition
     \\ $m$ $\gets$ pick a smaller value
     \\ $B$.add\_precondtion ($0 \leq x$ $\wedge$ $x < m$)
     \\ verify
     }
\end{algorithm}

\noindent The algorithm starts by assigning to $m$ the value of $x$ in the initial model, then iteratively decreases $m$. 
At each iteration, it adds a new precondition to $B$ to reduce the the range of $x$ and performs verification based on the updated $B$, to check whether there still exists a model with a smaller value of $x$ within the interval [$0,\ m$).

Picking a smaller value for $m$ (line 9) can be implemented either by sequentially decreasing the value or using a binary reduction (as in binary search) for acceleration. 
The current implementation first checks whether $x$ can be 0; if yes, the minimization stops as it has found the minimum; otherwise, it continues applying binary reduction. For more flexibility, it uses two user-specified parameters controlling termination:
\begin{itemize}
\item \emph{Tolerance}: lower bound on the size of interval used in the binary search algorithm;
\item \emph{Max iteration}: maximum number of verification iterations allowed when minimizing an integer.
\end{itemize} 

\noindent We have not endeavored to prove the correctness of the algorithms since the correctness of the approach (the ``proof of the pudding'') is embodied in the result: as the overall goal is to obtain a counterexample, the final criterion is whether the minimized value is still a counterexample, as established rigorously by the underlying proof technology. If not, the original unminimized counterexample still applies. 

\section{Experiment}
\label{sec: experiment}
\noindent A preliminary evaluation of the usability of the CEAM approach covers over 40 program versions\footnote{https://github.com/huangl223/Proof2Test/tree/main/examples} of 9 examples, including some adapted from software verification competitions \cite{weide2008incremental, bormer2011cost, klebanov20111st}. The examples (listed in Table \ref{table: experiment}) include: 
1) \e{ACCOUNT} introduced in Fig. \ref{listing: ACCOUNT};
2) a \e{CLOCK} class implementing a clock counting seconds, minutes, and hours;
3) a \e{HEATER} class implementing a heater adjusting it state (on or off) based on the current temperature and a user-defined temperature; 
4) a \e{LAMP} class describing a lamp equipped with a switch (for switching on/off the lamp) and a dimmer (for adjusting the light intensity of the lamp); 
5)a  \e{BINARY\_SEARCH} class implementing the binary search algorithm;
6) a \e{LINEAR\_SEARCH} implementing the linear search algorithm;
7) a \e{SQUARE\_ROOT} that calculates two approximate square
roots of a positive integer;
8) \e{MAX} from Fig. \ref{listing: max};
9) a \e{SUM\_AND\_MAX} class computing the maximum and sum of the elements of an array.

Each row in the table reports on the experiment result of a single example, which consists of multiple versions. Each version was intentionally injected with a fault, such as confusions between $+$ and $-$, $\leq$ and $<$, $>$ and $\geq$, missing loop invariant(s), pre- or postcondition, etc.
The experiments use AutoProof to verify the programs, produce counterexamples for all occurring proof failures, and minimize them. The \emph{tolerance} and \emph{max iteration} parameters are currently set to 0 and 20 respectively. 

The third column gives the total number of integer variables whose minimized in the experiment. Cases where no minimization is performed (e.g., the value of an integer variable in the counterexample is already 0 before minimization) are not included. The reduction rate (fourth column), number of iterations (fifth column), verification time (sixth column) and minimization time (last column) per integer are averaged out over all minimized integers of each example.

As the experiment result shows, CEAM minimization is cost-effective overall: in most cases, conducting minimization reduces the values of integer variables by over 80\% with an average cost of less than 4 extra verification runs (iterations). Most of the minimized integer values are fairly small and easy-to-understand: out of 125 minimized values, 108 are in the range [-2, 2], out of which 58 are zero; values not in that range are usually close to the values of some predefined constants in the program.


\begin{table*}[htbp]
   \caption{Experiment Results}
  \centering
   \renewcommand\arraystretch{1.3}
    \begin{tabular}{|c|c|c|c|c|c|c|}
    \hline
    Example  & \tabincell{c}{Number\\ of versions} & \tabincell{c}{Total Number of\\ Minimized Integers} &
     \tabincell{c}{Avg. Reduction\\ Rate} & \tabincell{c}{Avg. Number \\of Iterations} &
     \tabincell{c}{Avg. Verification \\Time (seconds)} & \tabincell{c}{Avg. Minimization\\ Time (seconds)} 
    \\ \hline
         ACCOUNT & 7 & 17 & 99.98\% & 2.5  & 0.028 & 0.087 
    \\ \hline
             CLOCK & 6 & 13  & 100\% & 1.46 & 0.019 & 0.034
    \\ \hline
             HEATER & 2 & 4 & 48.4\% & 4.25 & 0.030 & 0.128 
    \\ \hline
             LAMP &4 &8 & 0.819\% & 1.875 & 0.115 & 0.233
    \\ \hline
             BINARY\_SEARCH & 5& 31&98.8\% & 3.22&0.448 &1.512
        \\ \hline
            LINEAR\_SEARCH & 3& 9& 99.9\% & 3.44& 0.087& 0.279
    \\ \hline
            SQUARE\_ROOT & 4 & 3 & 89.9\% & 4& 0.133& 0.505
    \\ \hline
                MAX & 4 & 12 & 87.1\% & 4.25 & 0.213 & 1.456
    \\ \hline
             SUM\_AND\_MAX & 6 & 11& 80.7\%& 3.45& 0.590&1.704
    \\ \hline
    \end{tabular}%
  \label{table: experiment}%
\end{table*}%

\vspace{-0.05in}
\section{Related Work}
\label{sec: related_work}
\noindent In line with the objective of helping programmers to understand the causes of proof failures, several approaches have been proposed to provide more user-friendly visualizations of counterexample models \cite{le2011boogie, chakarov2022better, hauzar2016counterexamples, stoll2019smt, nilizadeh2022generating}.
Claire et al. \cite{le2011boogie} developed the Boogie Verification Debugger (BVD), which interprets a counterexample model as a static execution trace (i.e., a sequence of abstract states). David et al. \cite{hauzar2016counterexamples} transformed the models back into a counterexample trace comprehensible at the original source code level (SPARK) and display the trace using comments. Similarly, Aleksandar et al. \cite{chakarov2022better} transformed SMT models to a format close to the Dafny syntax. In contrast to the present work, these approaches concentrate on the generation of human-readable counterexample and do not consider any counterexample minimization.

Another direction of work to ease the understanding of proof failures is to generate more useful counterexamples in the first place: Polikarpova et al. \cite{polikarpova2013run} developed a tool, Boogaloo, which applies symbolic execution to generate counterexamples for failed Boogie programs. Like the present approach, Boogaloo displays minimal counterexamples in the form of valuations of relevant variables.  
M\"{u}ller et al. \cite{muller2011using} implemented a Visual Studio dynamic debugger plug-in for Spec\#, to reproduce a failing execution from the view of the prover. Likewise, Petiot et al. \cite{petiot2018testing, petiot2016your} developed STADY, which produces failing tests for the failed assertions using symbolic execution techniques. That approach is also referred to as \textit{testing-based counterexample synthesis}: it first translates the original C program into programs suitable for testing (run-time assertion checking), then applies symbolic execution to generate counterexamples (input for failing tests) based on the translated program. Unlike to that approach, CEAM counterexample extraction directly exploits the counterexample models produced by the provers, and hence does not require additional program instrumentation or counterexample generation.

\vspace{-0.1in}
\section{Conclusion}
\label{sec: conclusion}
\noindent This article has presented Counterexample Extraction and Minimization (CEAM), an approach that improves the quality of counterexamples generated in the presence of failed program proofs.
CEAM  automatically generates simple and easy-to-understand counterexamples.
We believe this makes the results of failed proofs practical enough to be used by regular programmers. 
The CEAM implementation is integrated in the AutoProof verifier to assist programmers when debugging failed proofs. 
The approach could also be applied to other Hoare-style verification tools relying on Boogie-style provers and SMT solvers. 

Ongoing work includes implementing a feature of automatic test generation based on the counterexamples produced by CEAM, as well as extending the scope of CEAM to include the supports for more data types and program constructs. We also plan to conduct systematic empirical studies to evaluate precisely the benefits of the proposed techniques for programmers with no verification expertise.

\textit{Acknowledgments:} We thank the anonymous referees for comments which led to significant improvements. The work benefitted from discussions with Filipp Mikoian, Alexander Kogtenkov and Alexander Naumchev from SIT.

\vspace{-0.1in}
\bibliographystyle{splncs04}
\bibliography{reference}

\end{document}